# Trivial topology-entailed states at interfaces between identical topological phases

**Authors**

Julius Beck[1]†, Francesco S. Piccioli[1,2]†, Mark Kremer[1], Max Ehrhardt[1], Lukas J. Maczewsky[1], Nora Timm[3], Matthias Heinrich[1], Iacopo Carusotto[2], Alexander Szameit[1*]

**Affiliations**

[1] Institute of Physics, University of Rostock, Albert-Einstein-Str. 23, 18059 Rostock, Germany.
[2] INO-CNR BEC Center and Dipartimento di Fisica, Università di Trento, Via Sommarive 14, 38123 Povo, Italy.
[3] Fraunhofer Research Institution for IMTE, Mönkhofer Weg 239a, 23562 Lübeck, Germany

*Corresponding author: alexander.szameit@uni-rostock.de
†These authors contributed equally to this work.

**Abstract**

Topological edge states arise at boundaries between bulk phases with distinct topological invariants and are characterized by robust, often chiral transport. When two domains sharing identical bulk topology are brought into contact, their counterpropagating edge modes are generally expected to hybridize and cancel each other out in the process. Here, we show that this cancellation can be lifted by trivial parameter variations at the interface, giving rise to a distinct class of gap-localized inter-face modes, which we term trivial topology-entailed (TTE) states. Using anomalous Floquet photonic lattices as a concrete platform, we experimentally observe the formation of bidirectional TTE states between topologically equivalent domains. In particular, we demonstrate their spectral confinement within bulk gaps, their coherent coupling to conventional chiral edge channels, and their ability to mediate interface-guided transport across composite lattices. Our results establish TTE states as a general interface phenomenon in composite topological systems and provide a unified framework for understanding hybridized boundary modes beyond conventional bulk-boundary correspondence.

## Introduction

Topological phases of matter are characterized by bulk invariants that dictate the existence of boundary states at interfaces between distinct topological domains.(*1*, *2*) This bulk–boundary correspondence underlies the emergence of robust edge transport in a wide range of systems, from electronic materials to photonic lattices.(*3–5*) When two domains with different topological indices are brought into contact, interface modes are expected to appear in a predictable manner, as illustrated schematically in Fig. 1a. Such boundary states constitute one of the central paradigms of topological physics.

However, the situation changes when two systems with identical bulk topology are brought into contact. In this case, each domain supports a counterpropagating edge mode at the interface, and the conventional expectation is that these modes destructively interfere and fully hybridize into the bulk, leaving no residual state in the gap, as illustrated in Fig. 1b. Since the bulk invariants on both sides are equal, bulk–boundary correspondence does not mandate the existence of an interface mode, and the interface is therefore expected to remain spectrally trivial.

This expectation implicitly assumes that the cancellation between the two counterpropagating edge modes is complete. However, perfect destructive interference requires strict symmetry between the two domains. If a trivial parameter of one system is modified - without altering its bulk topological index - the two edge modes no longer fully cancel. Instead, they hybridize into a pair of interface states that remain within the bulk gap, as illustrated in Fig. 1c. We refer to this class of interface modes as trivial topology entailed (TTE) states.

Boundary states induced by deliberately engineered defects in topological lattices - such as line-defect waveguides - have been reported in various photonic implementations.(*5–7*) In retrospect, such defect-induced states can be understood within the framework introduced here, as instances where the symmetry between counterpropagating boundary modes is lifted. However, these realizations have typically been interpreted in an implementation-specific manner. At the same time, other classes of interface modes - such as those observed in valley-Hall platforms - originate from distinct symmetry or valley-dependent topological mechanisms rather than from identical bulk topology.(*8*, *9*) The TTE framework introduced here therefore identifies and unifies a distinct class of interface states that arise specifically between topologically identical systems.

Here, we experimentally realize and characterize TTE states in a photonic lattice platform based on femtosecond laser-written waveguide arrays, a well-established setting for implementing topological band structures and Floquet topological phases.(*10–14*) By introducing a controlled parameter asymmetry between two topologically identical domains, we engineer interface bands that remain within the bulk gap despite identical bulk topology. We probe their properties through injection and extraction experiments, demonstrating coherent population transfer between TTE states and conventional topological edge modes. In addition, we investigate transport along the TTE channel and contrast it with bulk-mediated evanescent coupling, thereby elucidating the distinct interface character of these states.

## Results

**Concept and band formation of TTE states:** Anomalous Floquet topological insulators constitute a well-established class of periodically driven systems that have been theoretically(*15*) formulated and experimentally realized across multiple platforms, including photonic waveguide arrays(*11*, *16*) and ultracold atomic systems(*17*). More broadly, Floquet engineering has enabled the realization of a wide range of topological phases, including symmetry-protected(*18*), non-Hermitian(*19*) and nonlinear(*20*, *21*) extensions, underscoring the high degree of control and understanding achieved in driven systems. While the mechanism underlying TTE states is not restricted to periodically driven lattices, the Floquet platform offers a convenient experimental framework in which multiple spectral gaps and edge channels can be implemented within a single structure.

As a concrete implementation, we consider a honeycomb lattice subject to a three-step cyclic coupling protocol.(*22*) During one driving period $T$, which defines the Floquet period and the associated quasi-energy Brillouin zone, nearest-neighbor couplings are activated sequentially along the three bond directions of the lattice (Fig. 2a). The explicit time-periodic Hamiltonian and the calculation of the associated winding numbers are provided in the Supplementary Information.

The topological phase of this driven system is governed by the dimensionless coupling angle $\theta_c = CT/3$, where $C$ denotes the nearest-neighbor coupling strength. For $\pi/3 < \theta_c < 2\pi/3$, the lattice enters the anomalous Floquet regime characterized by chiral edge states traversing both the quasi-energy gap around $\varepsilon = 0$ and the gap at the Floquet zone boundary $\varepsilon = \pm\pi/T$ (Fig. 2b). In this system, the availability of topological edge modes in multiple spectral gaps provides a convenient route to realize TTE states, as it enables spectral overlap between boundary states of two domains once a rigid energy shift is introduced.

We now consider two adjacent domains governed by the same driving protocol. If the two domains are identical, the counterpropagating edge modes at their common interface fully hybridize and cancel, so that the composite system behaves as a single enlarged topological domain with a chiral channel only along its outer boundary.

To lift this cancellation without altering the bulk topological index, we introduce a uniform on-site detuning $\delta$ in one of the domains, which rigidly shifts its quasi-energy spectrum. In the illustrative case $\delta = \pi/T$, the quasi-energy gaps are effectively interchanged, so that edge modes originally residing in different gaps become energetically aligned across the interface. Instead of annihilating, the counterpropagating interface modes hybridize into a pair of bidirectional interface bands confined within the bulk gaps (Fig. 2c).

Importantly, this mechanism is not limited to the specific value $\delta = \pi/T$. As shown in Fig. 2d, TTE states emerge for any finite detuning $\delta \neq 0$, with their group velocity continuously tunable by the magnitude of the perturbation. In the regime of small $\delta$, the near-cancellation of the interface modes leads to a quasi-flat TTE band and correspondingly slow propagation along the interface.

**Experimental observation and coupling behavior:** We experimentally investigate TTE states in pairs of detuned honeycomb Floquet lattices realized by femtosecond laser-written waveguides in fused silica(*23*). The samples consist of two adjacent domains implementing the three-step driving protocol described above, with a controlled on-site detuning applied to one half of the lattice. The total propagation length corresponds to several Floquet periods, enabling direct observation of edge and interface transport.

We first demonstrate the injection of light from a conventional topological edge channel into the TTE interface state (Fig. 3a). Excitation of a single waveguide along the outer perimeter launches a chiral edge wave packet propagating clockwise. Upon reaching the interface between the two domains, a distinct portion of the optical intensity is transferred into the TTE channel and subsequently propagates downward along the interface. The experimentally recorded intensity distribution (top panel) is well reproduced by tight-binding simulations (bottom panel), confirming the interface-mediated routing of light.

Conversely, when light is launched into a waveguide located at the interface, a superposition of TTE modes is excited. These bidirectional interface states propagate both upward and downward and subsequently couple into the outer chiral edge channels at the respective boundaries (extraction, Fig. 3b). The observed population transfer between interface and perimeter modes directly demonstrates spectral overlap and coherent coupling between TTE states and conventional topological edge channels.

In both configurations, a fraction of the optical signal spreads into the bulk. This behavior is expected, as single-site excitation inevitably populates a broad range of quasi-momenta and therefore excites not only interface and edge bands but also bulk states. Nevertheless, the dominant intensity remains localized at the interface or outer boundary, consistent with the presence of well-defined gap states. The agreement between experiment and numerical simulations confirms that the bulk background does not obscure the distinct interface-mediated transport pathways.

Together, these observations establish the bidirectional character of TTE states, their confinement to the spectral gap, and their coherent coupling to chiral edge channels in adjacent domains.

**Interface-mediated transport and parameter dependence:** To further characterize the transport properties of TTE states, we implement an L-shaped geometry in which the interface channel connects topological edge states located at opposite sides of the composite lattice (Fig. 4a). In this configuration, light launched into the chiral edge channel at the upper boundary can couple into the TTE state, propagate along the interface over a distance $d$, and subsequently couple back into the edge channel at the lower boundary. The lateral size of the smaller domain is chosen such that the total propagation length precludes light from fully circumnavigating the outer perimeter, ensuring that any intensity detected at the lower edge must be mediated by the interface.

Experimentally, we quantify the transmittivity $T(d)$ as the fraction of optical intensity detected along the lower edge after excitation at the upper boundary. The measured data (Fig. 4b) reveal an initial decrease of $T(d)$ with increasing interface length, followed by a tendency towards a finite asymptotic value for sufficiently strong detuning. This behavior indicates that the transmission does not vanish for large interface lengths but instead approaches a non-zero plateau.

To interpret this observation, we compare the measurements to tight-binding simulations of both an idealized lattice and a model incorporating experimentally fitted parameters. The ideal system exhibits a clear saturation of $T(d)$ at large distances, reflecting coherent transport along a well-defined interface band. The realistic model reproduces the experimentally observed trend, including the gradual onset of the plateau. The agreement between experiment and simulation confirms that the observed transmission is mediated by the TTE band rather than by residual bulk coupling.

The dependence on the detuning strength further clarifies this mechanism. For larger values of $\delta$, the TTE band acquires a higher group velocity, leading to more efficient interface-mediated transmission within the finite propagation length of the sample. As $\delta$ decreases, the TTE band becomes increasingly flat, corresponding to a reduced group velocity. In this regime, finite propagation time limits the observable transmission, and the plateau signature is progressively suppressed. This continuous behavior is consistent with the band-structure picture introduced in Sec. 2.1 and does not rely on any special choice of detuning.

## Discussion

In this work, we have introduced trivial topology-entailed (TTE) states as a general class of interface modes that emerge between domains sharing identical bulk topological invariants when the cancellation of counterpropagating edge states is incomplete. Rather than relying on a mismatch of topological indices, the formation of TTE states is governed by the controlled lifting of destructive interference through trivial parameter variations across the interface. This perspective provides a unified framework for understanding interface-localized modes that arise between topologically equivalent systems.

While related interface states have appeared in specific settings—such as line-defect geometries or symmetry-engineered boundaries—the present work identifies their common physical origin in the structured hybridization of boundary modes. In this sense, TTE states do not constitute a new topological phase but represent a distinct type of gap-localized interface band whose existence is enabled, yet not protected, by the surrounding bulk topology.

Using a Floquet honeycomb lattice as a concrete realization, we experimentally demonstrated the bidirectional character of TTE states, their confinement within spectral gaps, and their coherent coupling to conventional chiral edge channels. The interface-mediated transmission measurements further show that these states sustain band-like transport over extended distances and exhibit continuously tunable group velocities. Importantly, the underlying mechanism does not depend on periodic driving and can, in principle, be implemented in static systems supporting multiple topological gaps.(*24*)

More broadly, the controlled formation of bidirectional interface bands between topologically equivalent domains offers a new avenue for investigating structured hybridization phenomena in composite topological systems. The tunability of the interface dispersion suggests opportunities to explore slow-propagation regimes within topological lattices, as well as controlled coupling between spatially separated edge channels. (*25*, *26*) Beyond the specific photonic platform considered here, the concept of TTE states establishes a systematic framework for engineering internal interfaces in driven, symmetry-protected, or higher-order topological systems, where the interplay between bulk topology and trivial perturbations can give rise to rich boundary physics.(*27–29*)

## Materials and Methods

### Tight-binding model and topological invariants:

The system is described by a time-periodic tight-binding Hamiltonian defined on a honeycomb lattice with two sublattices. One driving period $T$ consists of three piecewise constant steps of equal duration $T/3$, during which nearest-neighbor couplings of magnitude $C$ are activated sequentially along one of the three bond directions, while the remaining couplings are suppressed. In momentum space, the Hamiltonian during the $i$-th driving step reads

$$H_i(k) = \begin{pmatrix} 0 & Ce^{j\boldsymbol{k}\cdot\boldsymbol{b}_i} \\ Ce^{-j\boldsymbol{k}\cdot\boldsymbol{b}_i} & 0 \end{pmatrix}$$

where $\boldsymbol{b}_i$ denotes the nearest-neighbor bond vector associated with the active coupling direction. The full time-dependent Hamiltonian is given by

$$H(\boldsymbol{k},t) = \sum_{i=1}^{3} \mathbb{I}_{T_i}(t) H_i(k).$$

with $\mathbb{I}_{T_i}(t)$ the indicator function for the respective time interval.

A uniform on-site detuning $\delta$ is introduced in one of the two domains, leading to a rigid quasi-energy shift of its spectrum.

The dimensionless coupling angle is defined as $\theta_c = CT/3$. For $\pi/3 < \theta_c < 2\pi/3$, the lattice resides in the anomalous Floquet topological phase supporting chiral edge states in both quasi-energy gaps.

The topological properties are quantified by the (2+1)D winding number computed from the time-evolution operator over one driving period. The winding number at quasi-energy $\varepsilon$ is defined as

$$\mathcal{W}_\epsilon = \frac{1}{8\pi^2} \int \mathrm{Tr}\left(U_\epsilon^{-1} \partial U_\epsilon \left[U_\epsilon^{-1} \partial_{k_x} U_\epsilon, U_\epsilon^{-1} \partial_{k_y} U_\epsilon\right]\right) \mathrm{d}t \mathrm{d}k_x \mathrm{d}k_y,$$

where $U_\varepsilon$ is constructed from the time-evolution operator using a branch cut that distinguishes the 0 and $\pi$ quasi-energy gaps(*15*).

**Numerical methods:** Band structures were calculated in ribbon geometries with periodic boundary conditions along the interface and open boundaries in the transverse direction. The quasi-energy spectrum was obtained from the Floquet time-evolution operator over one driving period. The maximal group velocity of the TTE states was determined from the maximum slope of the corresponding quasi-energy band.

Light propagation in finite lattices (Figs. 3 and 4) was simulated by numerically solving the tight-binding coupled-mode equations with single-site excitation as initial condition. Unless otherwise stated, lattice dimensions and parameters were chosen to match the experimental realizations. The transmission $T(d)$ was evaluated as the fraction of total intensity localized at the output edge after full propagation.

**Waveguide fabrication and characterization:** Our photonic implementation of TTE states was based on laser-written lattices of evanescently coupled waveguides.(*30*) Ultrashort laser pulses from a frequency doubled fiber amplifier system (Coherent Monaco) at a carrier wavelength of 517 nm were focused (50x objective, NA = 0.6) into the volume of 150 mm long fused silica samples (Corning 7980). The evanescent coupling rate between adjacent guides was controlled by modulating the distance between them, from a distance of 9 µm for the coupling regions to 30 µm for the regions without optical coupling. Cosine-shaped transitions were employed to connect the individual steps of the driving protocol into contiguous waveguide trajectories. In turn, the inscription speed served as process parameter to implement different propagation constants, i.e. the desired shift of the on-site potential. The waveguide arrangements were designed for an operating wavelength of 633 nm, and characterized by injecting light from a continuous-wave

Helium-Neon laser light into specific lattice sites and subsequently imaging the output onto a CCD camera.

**References**

1. M. Z. Hasan, C. L. Kane, *Colloquium* : Topological insulators. *Rev. Mod. Phys.* **82**, 3045–3067 (2010).

2. X.-L. Qi, S.-C. Zhang, Topological insulators and superconductors. *Rev. Mod. Phys.* **83**, 1057–1110 (2011).

3. A. B. Khanikaev, S. Hossein Mousavi, W.-K. Tse, M. Kargarian, A. H. MacDonald, G. Shvets, Photonic topological insulators. *Nature Mater* **12**, 233–239 (2013).

4. L. Lu, J. D. Joannopoulos, M. Soljačić, Topological photonics. *Nature Photon* **8**, 821–829 (2014).

5. T. Ozawa, H. M. Price, A. Amo, N. Goldman, M. Hafezi, L. Lu, M. C. Rechtsman, D. Schuster, J. Simon, O. Zilberberg, I. Carusotto, Topological photonics. *Rev. Mod. Phys.* **91**, 015006 (2019).

6. J. Chen, W. Liang, Z.-Y. Li, Strong coupling of topological edge states enabling group-dispersionless slow light in magneto-optical photonic crystals. *Phys. Rev. B* **99**, 014103 (2019).

7. Z. Jiang, Y. Gao, L. He, H. Song, J. Zhou, R. Zhu, Helical edge states of topological photonic crystals with line defects. *Appl. Opt., AO* **58**, 2294–2299 (2019).

8. H. Xue, Y. Yang, B. Zhang, Topological Valley Photonics: Physics and Device Applications. *Advanced Photonics Research* **2**, 2100013 (2021).

9. A. B. Khanikaev, G. Shvets, Two-dimensional topological photonics. *Nature Photon* **11**, 763–773 (2017).

10. M. C. Rechtsman, J. M. Zeuner, Y. Plotnik, Y. Lumer, D. Podolsky, F. Dreisow, S. Nolte, M. Segev, A. Szameit, Photonic Floquet topological insulators. *Nature* **496**, 196–200 (2013).

11. S. Mukherjee, A. Spracklen, M. Valiente, E. Andersson, P. Öhberg, N. Goldman, R. R. Thomson, Experimental observation of anomalous topological edge modes in a slowly driven photonic lattice. *Nat Commun* **8**, 13918 (2017).

12. S. Stützer, Y. Plotnik, Y. Lumer, P. Titum, N. H. Lindner, M. Segev, M. C. Rechtsman, A. Szameit, Photonic topological Anderson insulators. *Nature* **560**, 461–465 (2018).

13. E. Lustig, L. J. Maczewsky, J. Beck, T. Biesenthal, M. Heinrich, Z. Yang, Y. Plotnik, A. Szameit, M. Segev, Photonic topological insulator induced by a dislocation in three dimensions. *Nature* **609**, 931–935 (2022).

14. G. G. Pyrialakos, J. Beck, M. Heinrich, L. J. Maczewsky, N. V. Kantartzis, M. Khajavikhan, A. Szameit, D. N. Christodoulides, Bimorphic Floquet topological insulators. *Nat. Mater.* **21**, 634–639 (2022).

15. M. S. Rudner, N. H. Lindner, E. Berg, M. Levin, Anomalous Edge States and the Bulk-Edge Correspondence for Periodically Driven Two-Dimensional Systems. *Phys. Rev. X* **3**, 031005 (2013).

16. L. J. Maczewsky, J. M. Zeuner, S. Nolte, A. Szameit, Observation of photonic anomalous Floquet topological insulators. *Nat Commun* **8**, 13756 (2017).

17. K. Wintersperger, C. Braun, F. N. Ünal, A. Eckardt, M. D. Liberto, N. Goldman, I. Bloch, M. Aidelsburger, Realization of an anomalous Floquet topological system with ultracold atoms. *Nat. Phys.* **16**, 1058–1063 (2020).

18. L. J. Maczewsky, B. Höckendorf, M. Kremer, T. Biesenthal, M. Heinrich, A. Alvermann, H. Fehske, A. Szameit, Fermionic time-reversal symmetry in a photonic topological insulator. *Nat. Mater.* **19**, 855–860 (2020).

19. A. Fritzsche, T. Biesenthal, L. J. Maczewsky, K. Becker, M. Ehrhardt, M. Heinrich, R. Thomale, Y. N. Joglekar, A. Szameit, Parity–time-symmetric photonic topological insulator. *Nat. Mater.*, 1–6 (2024).

20. L. J. Maczewsky, M. Heinrich, M. Kremer, S. K. Ivanov, M. Ehrhardt, F. Martinez, Y. V. Kartashov, V. V. Konotop, L. Torner, D. Bauer, A. Szameit, Nonlinearity-induced photonic topological insulator. *Science* **370**, 701–704 (2020).

21. S. Mukherjee, M. C. Rechtsman, Observation of Floquet solitons in a topological bandgap. *Science* **368**, 856–859 (2020).

22. T. Kitagawa, E. Berg, M. Rudner, E. Demler, Topological characterization of periodically driven quantum systems. *Phys. Rev. B* **82**, 235114 (2010).

23. A. Szameit, S. Nolte, Discrete optics in femtosecond-laser-written photonic structures. *J. Phys. B: At. Mol. Opt. Phys.* **43**, 163001 (2010).

24. F. Piccioli, "Topological Photonics in guided wave geometries: A topological toolbox for signal processing," thesis, Università degli studi di Trento (2023).

25. J. Guglielmon, M. C. Rechtsman, Broadband Topological Slow Light through Higher Momentum-Space Winding. *Phys. Rev. Lett.* **122**, 153904 (2019).

26. G. Arregui, J. Gomis-Bresco, C. M. Sotomayor-Torres, P. D. Garcia, Quantifying the Robustness of Topological Slow Light. *Phys. Rev. Lett.* **126**, 027403 (2021).

27. Z. Zhang, J. Wang, H. Qin, R. Fleury, Floquet topological physics in photonics [Invited]. *Opt. Mater. Express, OME* **15**, 828–889 (2025).

28. W. A. Benalcazar, B. A. Bernevig, T. L. Hughes, Quantized electric multipole insulators. *Science* **357**, 61–66 (2017).

29. F. Schindler, A. M. Cook, M. G. Vergniory, Z. Wang, S. S. P. Parkin, B. A. Bernevig, T. Neupert, Higher-order topological insulators. *Sci. Adv.* **4**, eaat0346 (2018).

30. A. Szameit, J. Burghoff, T. Pertsch, S. Nolte, A. Tünnermann, F. Lederer, Two-dimensional soliton in cubic fs laser written waveguide arrays in fused silica. *Opt. Express, OE* **14**, 6055–6062 (2006).

**Acknowledgments**

The authors would like to thank C. Otto for preparing the high-quality fused silica samples employed in this work.

**Funding:** AS received partial funding from European Research Council (grant EPIQUS), Deutsche Forschungsgemeinschaft (grants SCHE 612/6-1, SZ 276/12-1, BL 574/13-1, SZ 276/15-1 and SZ 276/20-1) and the Alfried Krupp von Bohlen and Halbach foundation. FSP acknowledges fundings from Deutscher Akademischer Austauschdienst (grant 57507869 and 57381332). IC and FSP acknowledge financial support from the European Union H2020-FETFLAG-2018-2020 project "PhoQuS" (n.820392) and from the Provincia Autonoma di Trento,partly through the Q@TN initiative. A.S. and M.H. acknowledge funding from the Deutsche Forschungsgemeinschaft via SFB 1477 'Light–Matter Interactions at Interfaces' (project no. 441234705).

**Author contributions:** FSP, NS and ME fabricated the photonic waveguide systems. FSP, MK, ME, LJM and MH conducted the measurements and evaluated the data. FSP, MK and IC developed the theoretical description. FSP and MK conducted the numerical simulations. JB and FSP developed the conceptual framework of the study and led the preparation of the manuscript. IC and AS supervised the work of their respective groups. All authors interpreted the results and contributed to the final version of the manuscript.

**Competing interests:** Authors declare that they have no competing interests.

**Data and materials availability:** The experimental data that support the findings of this study are available in an open-access data repository [reference follows].

## Figures

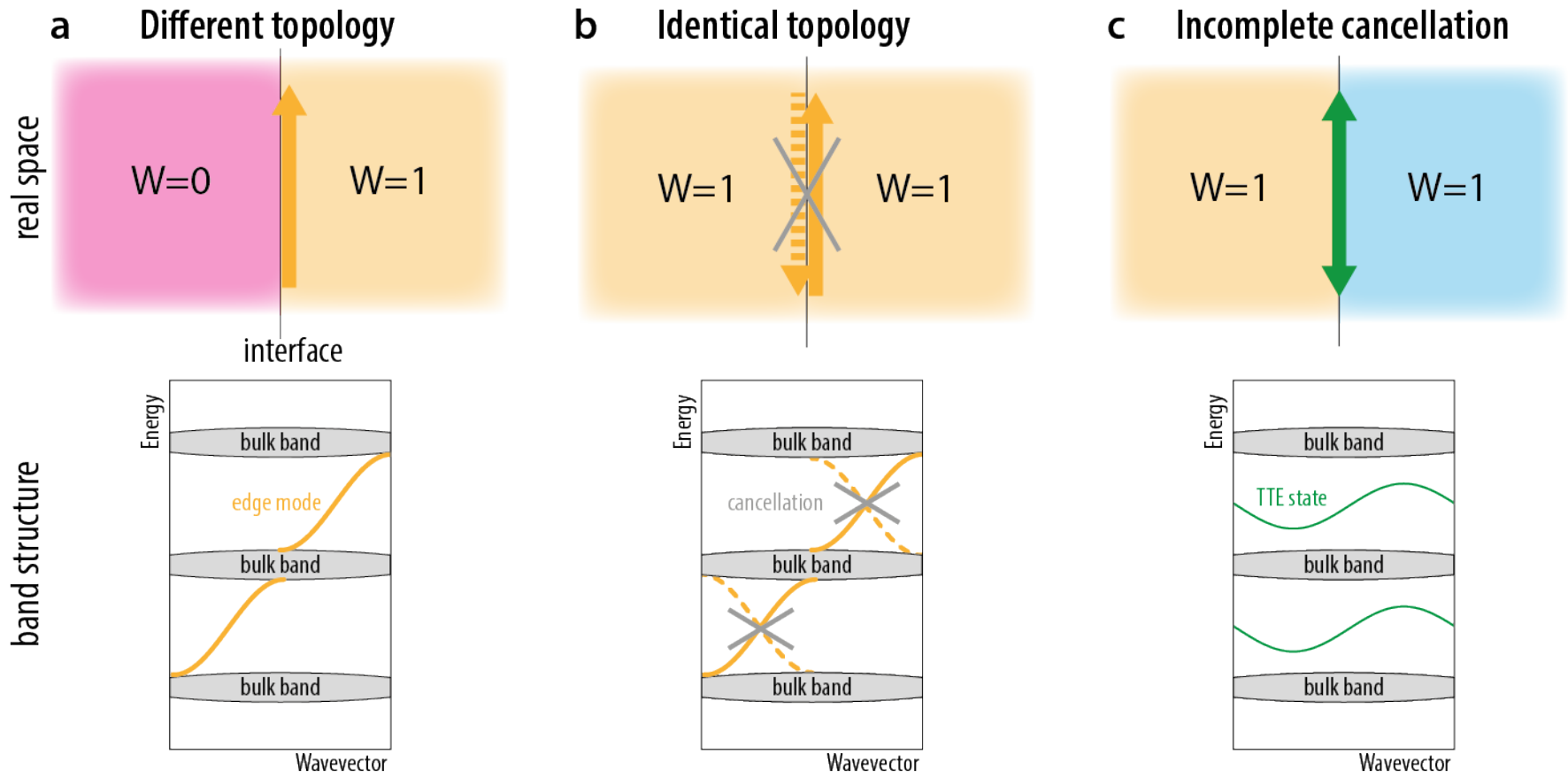

**Fig. 1. Conceptual illustration of trivial topology-entailed (TTE) states. (a)** Interface between domains with different topology. According to bulk–boundary correspondence, a chiral edge state emerges in the bulk bandgap. **(b)** Interface between domains with identical topological invariant $W$. The counter-propagating edge states localized at the interface hybridize and fully cancel, leaving the gap devoid of interface modes. **(c)** If the two domains differ by a trivial parameter (illustrated here by a detuning of one domain), this cancellation becomes incomplete. The hybridization of the counter-propagating edge states then gives rise to a pair of bidirectional interface modes, which we refer to as trivial topology-entailed (TTE) states. Top panels illustrate the interface geometry in real space, where the topological invariant $W$ of each domain is indicated. The bottom panels show schematic band structures highlighting the evolution of the interface states.

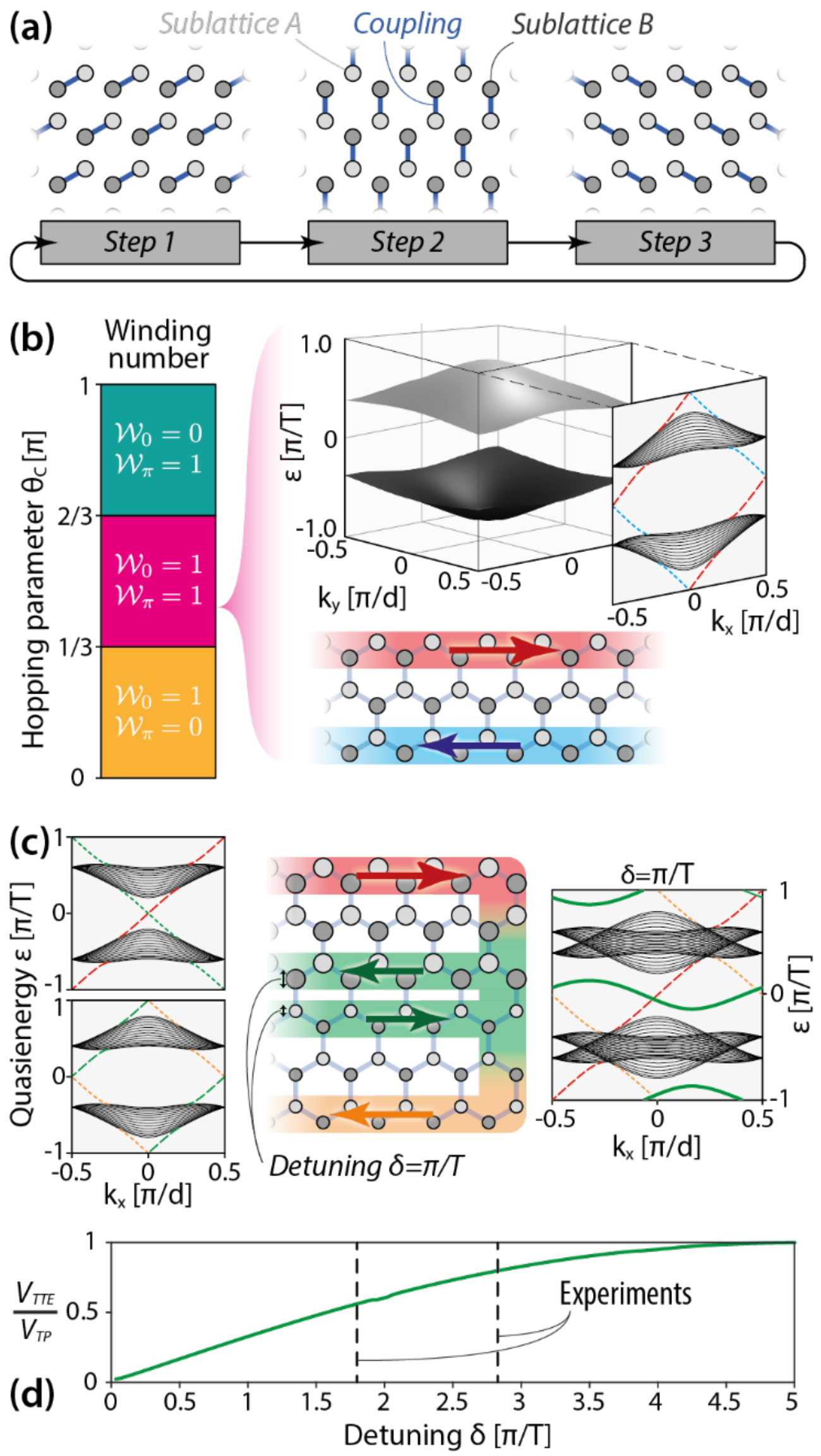

**Fig. 2. Formation of trivial topology-entailed (TTE) states. (a)** Three-step driving protocol of the periodically driven honeycomb lattice. In each step, nearest-neighbor coupling is activated exclusively along one of the three bond directions. **(b)** Left: Topological phase diagram as a function of the coupling phase $\theta_c = CT/3$. For $\pi/3 \leq \theta_c \leq 2\pi/3$, the system resides in the anomalous Floquet topological phase with winding numbers $W_0 = W_\pi = 1$. Right: Bulk quasi-energy spectrum (grey) and projected edge spectrum of a semi-infinite ribbon with zigzag termination along $x$ for $\theta_c = 0.4\pi$. Counterpropagating edge modes localized at opposite boundaries are highlighted. **(c)** Center: Composite system formed by two detuned but otherwise identical domains brought into contact along a zigzag interface. Left: Quasi-energy spectra of the isolated domains for $\delta = \pi/T$, shifted relative to each other. Right: Spectrum of the combined system. The counterpropagating interface modes hybridize into a pair of bidirectional TTE states (green). **(d)** Maximum group velocity of the TTE states relative to that of the chiral edge states as a function of detuning $\delta$ for $\theta_c = \pi/2$. The small dip near $\delta = 2$ arises from numerical overlap between bulk and interface bands, affecting their separation in the computation.

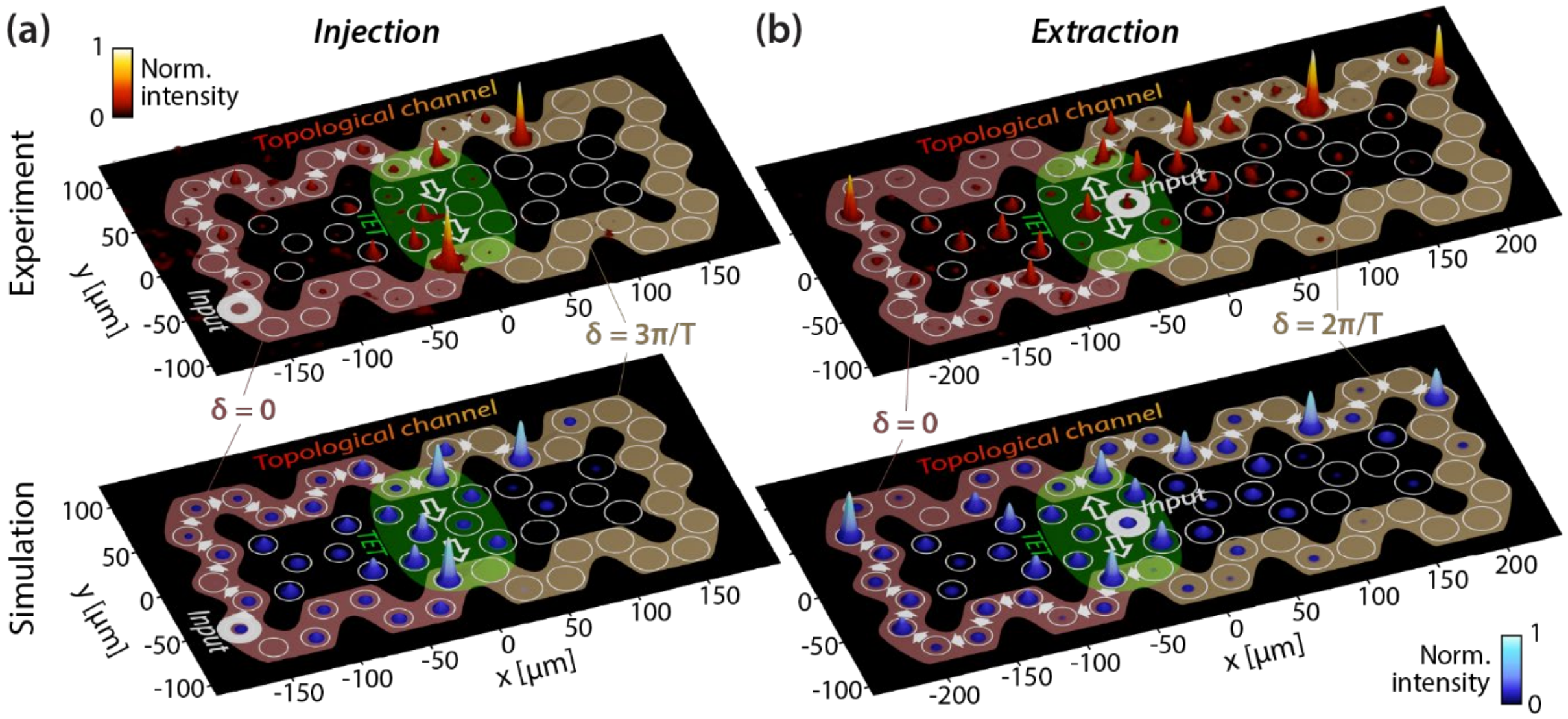

**Fig. 3. Bidirectional coupling between TTE and topological edge states.** Qualitative demonstration of injection **(a)** and extraction **(b)** of light between a chiral topological edge mode and a TTE state. In both cases, excitation is performed on a single lattice site and the output intensity distribution is recorded after propagation over 7 (left) and 8 (right) driving periods $T$, respectively. The two structures feature different interface detunings (indicated in the figure). Top: Measured output intensity distribution at the facet of the waveguide lattices.
Bottom: Corresponding tight-binding simulations. Due to the single-waveguide excitation used here, approximately 10–25% of the optical power overlaps with bulk modes, while the remaining intensity populates the interface channel. This bulk contribution is expected because a localized excitation does not select a single quasi-momentum or band and therefore has finite overlap with several Bloch modes. Minor apparent distortions of individual intensity maxima arise from the imaging conditions; all waveguides operate strictly in the single-mode regime.

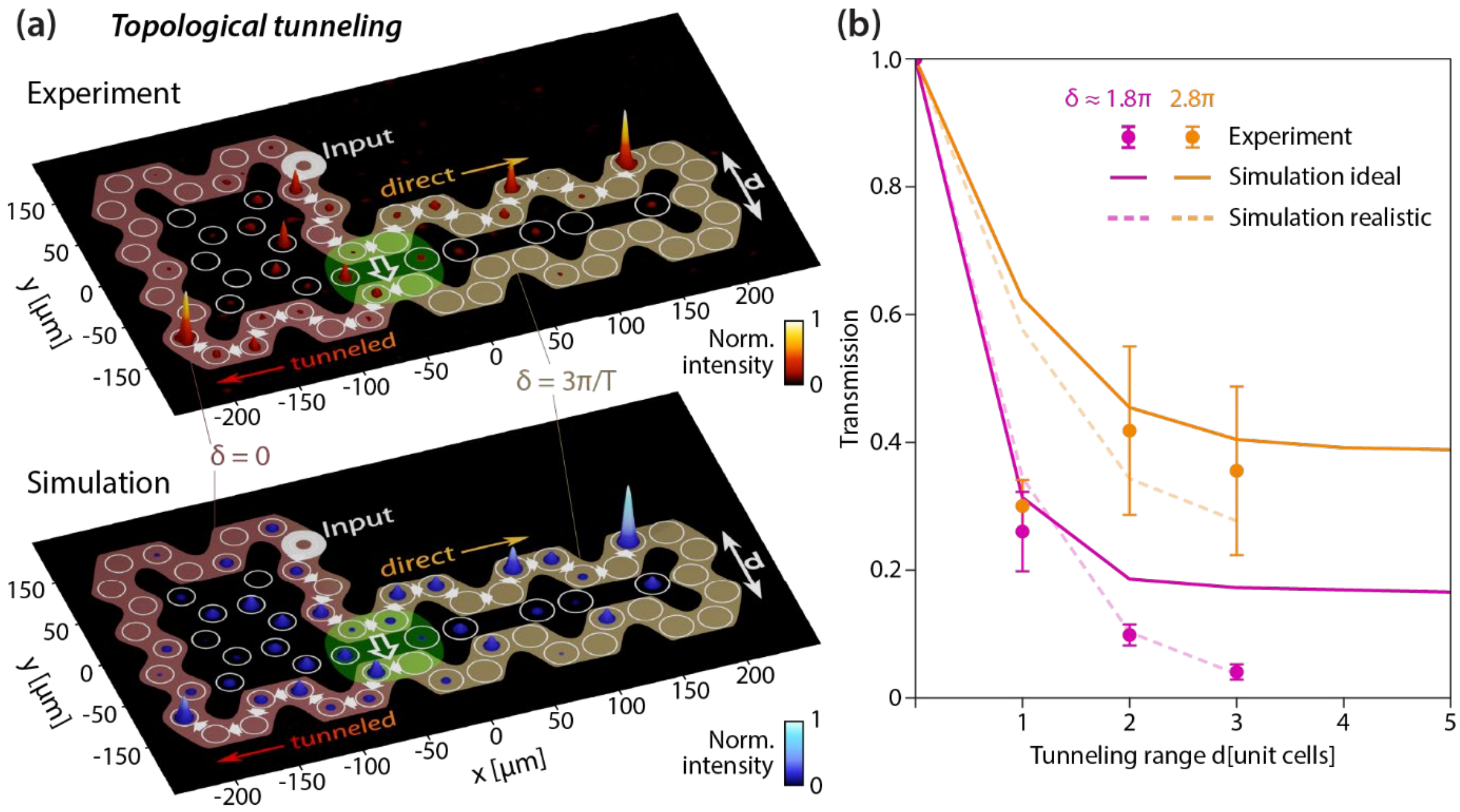

**Fig. 4. Interface-mediated transmission via TTE states. (a)** Measured (top) and simulated (bottom) output intensity distribution for an interface length $d = 2$ unit cells after propagation over seven driving periods. Light injected at the upper edge is transmitted to the lower edge through the TTE interface channel. **(b)** Transmission $T$ as a function of the interface length $d$. Experimental results are compared to tight-binding simulations for an ideal system ($\theta_c = 0.5\pi$, calculated over an extended range of $d$) and for a realistic system ($\theta_c = 0.4\pi$) using parameters extracted from the measurements. The numerical results indicate saturation towards a finite value $T > 0$ at larger interface lengths. In the realistic system, this saturation is not observed within the experimentally accessible propagation length due to finite propagation time.